# Random Shuffling to Reduce Disorder in Adaptive Sorting Scheme

Md. Enamul Karim and Abdun Naser Mahmood
AI and Algorithm Research Lab
Department of Computer Science, University of Dhaka
enam@du.bangla.net

***Abstract***

*In this paper we present a random shuffling scheme to apply with adaptive sorting algorithms. Adaptive sorting algorithms utilize the presortedness present in a given sequence. We have probabilistically increased the amount of presortedness present in a sequence by using a random shuffling technique that requires little computation. Theoretical analysis suggests that the proposed scheme can improve the performance of adaptive sorting. Experimental results show that it significantly reduces the amount of disorder present in a given sequence and improves the execution time of adaptive sorting algorithm as well.*

## 1. Introduction

Sorting is the process of arranging numbers in either increasing (non-decreasing) or decreasing (non-increasing) order. Due to its immense importance, sorting has never ceased to appear as research topic in computer science studies. The problem of sorting has been extensively studied and many important discoveries have been made in this respect.

It has been observed [Estivill-Castro and Wood 1991] that the performance of sorting algorithms can be increased if the presortedness of the data is taken into account. This led to a new type of sorting approach, which is now known as adaptive sorting.

Adaptive-sorting Algorithms take the advantage of existing order in the input. Intuitively 'run time' of an adaptive sorting algorithm varies smoothly from O(n) to O(n log n) as the disorder or "unsortedness" of the input varies [Mehlhorn 1984]. Adaptive sorting algorithms are important because nearly sorted sequences are common in practice [Knuth 1973, p.339; Sedgewick 1980, p.126; Mehlhorn 1984, p.54]. Adaptive Sorting Algorithm is not only theoretically interesting, but also it creates the possibility of improving upon existing sorting algorithms that do not take notice of existing order in the input data.

In this paper we have presented a randomization scheme to increase the presortedness of a given sequence so that performance of adaptive sorting improves. We use Step-Down-Runs as the measure of disorder and assume that all the elements in a given sequence have distinct values.

## 2. Previous Work

Earliest literature in this field is that of W.H Burge [Burge 1958]. In "*Sorting trees and measures of order*", he claims that sorting algorithms perform best if we take into account the pre-existing order of the data. He also proposed measures of disorder to evaluate the extent to which elements are already sorted.

Initial research on adaptive sorting took three different directions. First, during 1977-1980 there was a trend of research on insertion based internal sorting algorithms, which made use of specialized data structures that helped retain the sorted portion of the given sequence [Brown and Tarzan 1980; Guibas et al. 1977; Mehlhorn 1979]. Second, Cook and Kim [1980] conducted empirical studies, which resulted in Cook-Kim division and other partition methods. Cook-Kim division divides the input data into two parts such that one of the part is sorted and the other part is not. Using this division Cook-Kim constructed *Ck*sort, which splits X into two sequences, one of which is sorted and the other is sorted using a sorting algorithm called Quickersort. Third, Dijkstra [1982] introduced Smooth Sort, which uses a variation of Heap tree to implement the sorting algorithm.

Later, Mannila took into consideration all those efforts and introduced a formal model for the analysis of adaptive sorting algorithms in 1985 [Mannila 1985]. His model consists of the concept of presortedness as a function that evaluates disorder, and the definition of optimal adaptivity of a sorting algorithm with respect to a certain measure of disorder.

Following Mannila's work, researchers developed quite a few notable works that focused on the types of measure of disorders that an algorithm is adaptive to.

In 1988, Skiena published a sorting algorithm called *Melsort*[Skiena 1988]. Melsort constructs a partition of the input X, that consists of a set of sorted lists called the *encroaching lists* of X. In the final step Melsort, sorts the remaining unsorted lists and merges all lists to obtain the elements of X in sorted order.

In 1990, Levcopoulos and Petersson published *S*lab *S*ort [Levcopoulos and Petersson 1990]. Slab Sort achieves optimality with respect to a measure of disorder *SMS(X),* which ensures optimality with respect to measure of disorders *D*is, *M*ax, *R*uns, and *SUS*. Although slab sort is an important theoretical concept it has little practical value because it requires repeated median finding. The partitioning is similar to Quicksort, but with p-1 pivots it can be carried out in $O(|X|(1+\log[p+1]))$ time.

In 1991, Moffat and Petersson designed an insertion-based sorting algorithm, using a complex data structure called a *historical search tree* that makes an optimal number of comparisons with respect to a disorder *Reg*, but does not make an optimal number of data moves; it takes $\Omega(\log Reg(X) + |X|\log\log|X|)$ time [Moffat and Petersson 1991;1992]. However, this algorithm has size restrictions due to its large overhead

Estivill-Castro and Derick Wood conducted a series of work on adaptive sorting. Their work includes the development of generic adaptive sorting algorithm that focuses on the combinatorial properties of measures of disorder rather than the combinatorial properties of the algorithm.[Estivill-Castro and Wood 1991,1992]. The scheme is developed for both fixed and variable partitioning. Using them they obtained practical adaptive sorting algorithms optimal with respect to several important measures of disorder. They also introduced randomized generic sort, where partitions in a given sequence are created by a randomized division protocol. [Estivill-Castro and Wood 1992]

## 3. Measure of Disorder

We use Step-down *R*uns [Knuth 1973, p 161] as the measure of disorder. We determine the disorder for a given sequence in the following way.

### *Determination of M(X)*

In *Step-down Runs* the disorder is measured by the number of consecutive inverses. Specifically, disorder M(X) of a sequence X is measured as below.

Let $X = X_1, X_2, \ldots, X_{|X|}$ be a given permutation, then disorder of X, $M(X) = \sum_{i=1}^{|x|-1} Ci$

Where, $C_i = 1$ if $X_i > X_{i+1}$ and $C_j = 0$ if $X_i <= X_{i+1}$

It is easy to see that disorder ranges from 0 to |X|-1 for a sequence with |X| elements.

## 4. Random shuffle

By random shuffle we mean in this paper an arbitrary change in positions of two elements in an array. It is arbitrarily done to save computation. We also assume that in one shuffle two elements are swapped.

## 5. Observations

**Claim 1** Let M(X)= $z$ and $z'$ respectively, before and after a random shuffle on a sequence X and n = |X|. If z > $\lfloor n/2 \rfloor + 1$ then the probability $p_{<z}$ that, $z' < z$ is greater than .5 and $p_{<z}$ approaches 1, as z approaches n-1.

**Proof:**

Let, $\left\langle {n \atop k} \right\rangle$ be the number of permutations with M(X)= *k*, where $0 \le k \le n-1$.

For a sequence of $k = z$, the probability $p_{<z}$ that after a random shuffle *k* will be less than *z* is,

$$p_{<z} = \frac{\sum_{k=0}^{k<z} \left\langle {n \atop k} \right\rangle}{\sum_{k=0}^{k \le n-1} \left\langle {n \atop k} \right\rangle}$$

We have used Step-down *R*uns as the measure of disorder. We can relate $\left\langle {n \atop k} \right\rangle$ to Eulerian Numbers $\left\langle {n \atop k} \right\rangle_{eul}$, which represents number of permutations with *R*uns *k* [KNUTH, D. E. 1973, page 37], by the following relationship.

$$\left\langle {n \atop k} \right\rangle = \left\langle {n \atop k+1} \right\rangle_{eul}$$

So it can be written as,

$$p_{<z} = \frac{\sum_{k=1}^{k \le z} \sum_{0 \le j \le k} (-1)^j (k-j)^n \binom{n+1}{j}}{\sum_{k=1}^{k \le n} \sum_{0 \le j \le k} (-1)^j (k-j)^n \binom{n+1}{j}}$$

For odd *n*'s,

$$\left\langle {n \atop m-1} \right\rangle < \left\langle {n \atop m} \right\rangle \text{, for } 0 < m \le \lfloor n/2 \rfloor$$

$$\left\langle {n \atop m-1} \right\rangle > \left\langle {n \atop m} \right\rangle, \text{ for } \lfloor n/2 \rfloor < m \le n-1$$

and

$$\left\langle {n \atop \lfloor n/2 \rfloor - m} \right\rangle = \left\langle {n \atop \lfloor n/2 \rfloor + m} \right\rangle, \text{ for } 0 \le m \le \lfloor n/2 \rfloor$$

And for even *n*'s

$$\left\langle {n \atop m-1} \right\rangle < \left\langle {n \atop m} \right\rangle, \text{ for } 0 < m \le \lfloor n/2 \rfloor\text{-1 (floor is taken for symmetry)}$$

$$\left\langle {n \atop m-1} \right\rangle > \left\langle {n \atop m} \right\rangle, \text{ for } \lfloor n/2 \rfloor < m \le n-1$$

and $$\left\langle {n \atop \lfloor n/2 \rfloor - 1 - m} \right\rangle = \left\langle {n \atop \lfloor n/2 \rfloor + m} \right\rangle, \text{ for } 0 \le m \le \lfloor n/2 \rfloor\text{-1}$$

So we conclude that, the distribution is symmetric to a maximum value of disorder (table 1) and hence if z > $\lfloor n/2 \rfloor + 1$,

$$\sum_{k=0}^{k<z} \left\langle {n \atop k} \right\rangle > \sum_{k=z}^{k<n-1} \left\langle {n \atop k} \right\rangle$$

$$\therefore\ p_{<z} = \frac{\sum_{k=0}^{k<z} \left\langle {n \atop k} \right\rangle}{\sum_{k=0}^{k<z} \left\langle {n \atop k} \right\rangle + \sum_{k=z}^{k\le n-1} \left\langle {n \atop k} \right\rangle} > 0.5$$ and $p_{<z}$ approaches 1, as z approaches n-1.

**Table 1:** Disorders versus Number of Permutation for different |X|s

| | Disorders | | | | | | | |
|---|---|---|---|---|---|---|---|---|
| **\|X\|** | **0** | **1** | **2** | **3** | **4** | **5** | **6** | **7** |
| **1** | 1 | | | | | | | |
| **2** | 1 | 1 | | | | | | |
| **3** | 1 | 4 | 1 | | | | | |
| **4** | 1 | 11 | 11 | 1 | | | | |
| **5** | 1 | 26 | 66 | 26 | 1 | | | |
| **6** | 1 | 57 | 302 | 302 | 57 | 1 | | |
| **7** | 1 | 120 | 1191 | 2416 | 1191 | 120 | 1 | |
| **8** | 1 | 247 | 4293 | 15619 | 15619 | 4293 | 247 | 1 |

It should be noted that if we take a long array and divide it into parts, the disorders of the parts would also follow the above distribution.

**Claim 2** If there are $l$ parts (it is already stated that if an array is partitioned into number of parts, disorder pattern will follow the distribution shown in claim1) with $z > \lfloor n/2 \rfloor + 1$, then the probability that $c$ or more parts will have a $z_{new} < z$ after shuffling is,

$$\sum_{x=c}^{l} \binom{l}{x} p_{<z}^{l} (1 - p_{<z})^{l-x}$$

**Proof:**
From binomial distribution the claim follows.

## 6. Proposed Scheme

***Adaptive Sort with Random Shuffle (X)***

- Divide X into k subsequences
- Check the measure of disorder of each of k subsequences, if a subsequence has a disorder greater than a predetermined threshold $z$, apply *Random Shuffling Scheme* on that subsequence.
- Apply an adaptive-sorting algorithm to X

## 7. Analysis

We perform present analysis based upon generic adaptive sort proposed in [Estivill-castro and D Wood, 1992].

For $|X|$ elements and k parts there are $|X|$/k elements in each part. Let, shuffling be applied to parts with disorder greater than $z$ and there be at most $l$ *(l<=k)* such parts. Also assume that the number of shuffling applied in each of $l$ parts does not exceed ($|X|$/k)/m shuffles. Then, we conclude that the total cost of shuffling $C_{shuf}$ is,

$C_{shuf} \leq ((|X|/k)/m) * l = |X|/m$

By theorem 1.1 in [Estivill-castro and D Wood, 1992] we find that Generic Sort is adaptive to the measure M and it takes $O(|X|(1+\log[M(X)+1]))$ time in the worst case, where $|X|$ is the number of elements in *X*.

Let, after random shuffling, *M(X)* be $M(X)_{new}$. For simplicity, if we assume that the cost of shuffling is comparable to cost of comparsion then shuffling improves performance of the sorting scheme if,

$$O(|X|(1 + \log[M(X)_{new} + 1]) + C_{shuf}) < O(|X|(1 + \log[M(X) + 1]))$$

$$\Rightarrow \quad |X| \log[\frac{M(X)+1}{M(X)_{new}+1}] > C_{shuf}$$

$$\Rightarrow \quad |X| \log[\frac{M(X)+1}{M(X)_{new}+1}] > |X|/m$$

$$\Rightarrow \quad \frac{M(X)+1}{M(X)_{new}+1} > 2^{1/m}$$

Experimental results show that the condition can be satisfied.

## 8. Test Results

To produce test results we used adaptive merged sort algorithm as presented in [Estivill-castro and D Wood, 1992]. $k = 16$ has been considered. During random shuffling two random exchanges (m=2) have been performed in each subsequence with disorder $z \geq 10$. The result is summarized in table 2.

**Table 2: Comparative Results (time in comparative seconds)**

| Data Size | Disorder Before Shuffling | Disorder After Shuffling | Shuffling Time | Adaptive Sorting Time After Applying Shuffling | Shuffling + Adaptive Sorting Time | Adaptive Sorting Time Without Applying Shuffling | Non Adaptive Sorting Time | % of improvement w.r.t. Adaptive sorting Without Shuffling | % of improvement w.r.t. Non Adaptive Sorting |
|---|---|---|---|---|---|---|---|---|---|
| | 2503 | 2245 | 0.00044 | 0.0496 | 0.05004 | 0.05174 | 0.06546 | 3.28566 | 23.55637 |
| | 2501 | 2256 | 0.00042 | 0.04848 | 0.0489 | 0.04942 | 0.06524 | 1.05221 | 25.04598 |
| 5000 | 2492 | 2243 | 0.00048 | 0.0487 | 0.04918 | 0.05074 | 0.06284 | 3.0745 | 21.73775 |
| | 2502 | 2254 | 0.00038 | 0.04936 | 0.04974 | 0.049 | 0.06504 | -1.5102 | 23.52399 |
| | 2492 | 2248 | 0.00048 | 0.04886 | 0.04934 | 0.0512 | 0.06372 | 3.63281 | 22.56748 |
| | 5010 | 4506 | 0.0006 | 0.10564 | 0.10624 | 0.10996 | 0.1382 | 3.38305 | 23.1259 |
| | 5003 | 4508 | 0.00064 | 0.105 | 0.10564 | 0.10872 | 0.1394 | 2.83297 | 24.21808 |
| 10000 | 5009 | 4494 | 0.00064 | 0.10492 | 0.10556 | 0.10692 | 0.13812 | 1.27198 | 23.5737 |
| | 4991 | 4485 | 0.0062 | 0.105 | 0.1112 | 0.10876 | 0.13944 | -2.24347 | 20.25244 |
| | 5013 | 4497 | 0.0006 | 0.105 | 0.1056 | 0.10808 | 0.13812 | 2.2946 | 23.54474 |
| | 9994 | 8974 | 0.00112 | 0.19436 | 0.19548 | 0.19868 | 0.25624 | 1.61063 | 23.71214 |
| | 9992 | 8979 | 0.00152 | 0.19304 | 0.19456 | 0.1994 | 0.25564 | 2.42728 | 23.89297 |
| 20000 | 9995 | 8973 | 0.00116 | 0.19312 | 0.19428 | 0.20068 | 0.25436 | 3.18916 | 23.62007 |
| | 10019 | 8989 | 0.00166 | 0.19188 | 0.19354 | 0.198 | 0.2556 | 2.25253 | 24.28013 |
| | 9995 | 8987 | 0.00144 | 0.1926 | 0.19404 | 0.19992 | 0.25436 | 2.94118 | 23.71442 |
| | 15003 | 13461 | 0.00188 | 0.28744 | 0.28932 | 0.29936 | 0.38128 | 3.35382 | 24.11876 |
| | 15001 | 13468 | 0.00184 | 0.28752 | 0.28936 | 0.29808 | 0.38436 | 2.92539 | 24.71641 |
| 30000 | 14984 | 13462 | 0.00164 | 0.28748 | 0.28912 | 0.3 | 0.3856 | 3.62667 | 25.02075 |
| | 14988 | 13476 | 0.00176 | 0.28444 | 0.2862 | 0.29996 | 0.38316 | 4.58728 | 25.30536 |
| | 15014 | 13454 | 0.00192 | 0.28752 | 0.28944 | 0.29748 | 0.38504 | 2.7027 | 24.82859 |
| | 19984 | 18041 | 0.00236 | 0.3905 | 0.39286 | 0.4065 | 0.5155 | 3.35547 | 23.79049 |
| | 19966 | 18021 | 0.00232 | 0.3905 | 0.39282 | 0.3985 | 0.5 | 1.42535 | 21.436 |
| 40000 | 19966 | 17925 | 0.00268 | 0.39 | 0.39268 | 0.399 | 0.5 | 1.58396 | 21.464 |
| | 20036 | 17973 | 0.00236 | 0.3825 | 0.38486 | 0.4065 | 0.508 | 5.32349 | 24.24016 |
| | 19944 | 17955 | 0.00256 | 0.385 | 0.38756 | 0.3985 | 0.5075 | 2.74529 | 23.6335 |
| | 25016 | 22489 | 0.00324 | 0.4995 | 0.50274 | 0.508 | 0.6405 | 1.03543 | 21.5082 |
| | 25043 | 22476 | 0.00312 | 0.492 | 0.49512 | 0.516 | 0.6585 | 4.04651 | 24.81093 |
| 50000 | 24932 | 22490 | 0.00335 | 0.4995 | 0.50285 | 0.516 | 0.6405 | 2.54845 | 21.49102 |
| | 2501 | 22407 | 0.00298 | 0.5 | 0.50298 | 0.5235 | 0.6405 | 3.91977 | 21.47073 |
| | 25010 | 22407 | 0.00308 | 0.5 | 0.50308 | 0.5235 | 0.6405 | 3.90067 | 21.45511 |
| | 30055 | 26923 | 0.0042 | 0.594 | 0.5982 | 0.609 | 0.7815 | 1.7734 | 23.45489 |
| | 30041 | 26987 | 0.00396 | 0.578 | 0.58196 | 0.6095 | 0.782 | 4.51846 | 25.58056 |
| 60000 | 30096 | 26940 | 0.0042 | 0.578 | 0.5822 | 0.61 | 0.7735 | 4.55738 | 24.73174 |
| | 30003 | 26917 | 0.004 | 0.5935 | 0.5975 | 0.602 | 0.774 | 0.74751 | 22.80362 |
| | 30003 | 26911 | 0.00402 | 0.578 | 0.58202 | 0.602 | 0.765 | 3.31894 | 23.91895 |
| | 34958 | 31304 | 0.00488 | 0.687 | 0.69188 | 0.7035 | 0.914 | 1.65174 | 24.30197 |
| | 34943 | 31358 | 0.005 | 0.688 | 0.693 | 0.7105 | 0.914 | 2.46305 | 24.17943 |
| 70000 | 34979 | 31469 | 0.0046 | 0.672 | 0.6766 | 0.7105 | 0.922 | 4.77129 | 26.61605 |
| | 34958 | 31445 | 0.00442 | 0.679 | 0.68342 | 0.703 | 0.906 | 2.78521 | 24.56733 |
| | 34996 | 31486 | 0.00432 | 0.6795 | 0.68382 | 0.7035 | 0.9065 | 2.79744 | 24.56481 |
| | 39910 | 35858 | 0.00502 | 0.789 | 0.79402 | 0.8505 | 1.047 | 6.6408 | 24.16237 |
| | 39985 | 35916 | 0.00524 | 0.8045 | 0.80974 | 0.8365 | 1.047 | 3.19904 | 22.66094 |
| 80000 | 39932 | 35826 | 0.00532 | 0.8045 | 0.80982 | 0.8365 | 1.039 | 3.18948 | 22.05775 |
| | 40101 | 35933 | 0.00525 | 0.797 | 0.80225 | 0.82 | 1.039 | 2.16463 | 22.78633 |
| | 40035 | 35930 | 0.00516 | 0.805 | 0.81016 | 0.828 | 1.0395 | 2.15459 | 22.06253 |

## 9. Conclusion

The performance of the scheme in improving sorting time is quite impressive. Optimum values for k, z and m are left for future study. We considered only Step-down *R*uns as the measure of disorder. The idea of applying random shuffle to reduce other disorder measures should be studied.